\documentclass[11pt,twoside]{article}

\usepackage{asp2004}
\usepackage{psfig}
\usepackage{epsf}
\usepackage{graphics}
\usepackage{lscape}

\def\edcomment#1{\iffalse\marginpar{\raggedright\sl#1\/}\else\relax\fi}
\reversemarginpar

\begin{document}

\title{%
Search for a Shock Wave around the Crab Nebula}

\author{%
C.W.~Mauche, P.~Gorenstein}

\affil{%
Harvard-Smithsonian Center for Astrophysics, Cambridge, MA}

\begin{abstract}
We have searched the region surrounding the Crab Nebula for the existence of a
shock wave with the imaging instruments of the Einstein Observatory. The search is
complicated by the scattering of nebula and pulsar X-rays from the imperfectly
polished surfaces of the telescope mirror, as well as from interstellar grains
along the line of sight. Both of these effects lead to the appearance of X-ray
emission, in the form of an X-ray halo, beyond the boundaries of the nebula
filaments.

We show that the size, shape, and intensity of the halo around the Crab Nebula,
above the contribution of mirror scattering, is consistent with what is expected
from the scattering from interstellar grains. The upper limit on the X-ray emission
from a shock wave is about 1\% of the total 0.5--4 keV luminosity of the Crab or
about $2\times 10^{35}~\rm erg~s^{-1}$ (assuming a distance of 2.2 kpc). This
figure applies to a shell whose angular radius is 9 arcminutes. The upper limit is
smaller (larger) for a shell of larger (smaller) size. This upper limit is an order
of magnitude or more below the flux of  Cas A, Tycho, and Kepler SNRs, which are 2
to 3 times younger, but it is still above that of SN 1006.
\end{abstract}

\thispagestyle{plain}

\section{Introduction}

As a continuation of our study of the effects of scattering of X-rays by
interstellar grains, we have undertaken an examination of the surface brightness
profile (SBP) of the Crab Nebula. Toward this end, we have investigated the
large-scale structure of the nebula related to the existence of an X-ray halo.
Previous observations have offered convincing evidence for such a halo (Ku, 
et al.\ 1976; Toor, et al.\ 1976; Charles \& Culhane 1977), though a decisive
explanation as to its existence has, until the present, been lacking. Two
competing theories have been advanced. First, in analogy with other SNRs, the
halo is attributed to the emission of a strong shock. This interpretation is
troubled by the lack of X-ray emission lines from this region (Schattenburg, et
al.\ 1980), though it can not be excluded by this fact alone, given the
featureless X-ray spectrum for the SN 1006 remnant (Becker, et al.\ 1980). A
second theory which explains the X-ray halo is that it is due to the scattering
of X-rays by interstellar grains. We argue below that this interpretation is the
correct one, and discuss the Crab's halo as a natural, observable feature of all
bright galactic X-ray sources.

\section{The Measured Surface Brightness Profiles}

The Crab Nebula was imaged by both the High Resolution Imager (HRI) and the Imaging
Proportional Counter (IPC) aboard the Einstein Observatory. A quantitative display
of the HRI data is presented in Figure 1a. This is the azimuthally summed, radially
binned SBP. The binning neglects the ellipticity of the nebula, and is centered
roughly 16 arcseconds northwest of the pulsar at the centroid of the light in the
phase-averaged image. Three things are readily apparent: (1) the surface brightness
drops off dramatically beyond about 50 arcseconds, (2) a bump in this trend kicks
in at about $10^2$ arcseconds, and (3) we run out of the HRI's field of view just
when things are getting interesting.

\begin{figure}[!ht]
\plotfiddle{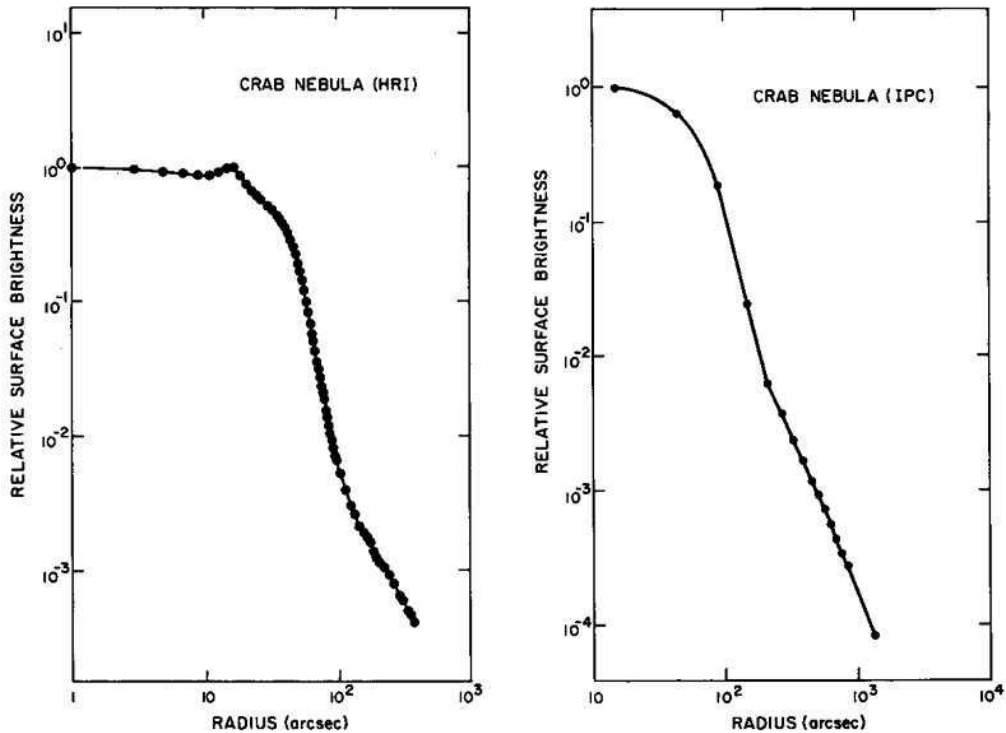}{3.99in}{0}{100}{100}{-194}{0}
\caption{The azimuthally summed, radially binned SBPs of the Crab Nebula. (a) The
HRI profile reveals the pulsar (at $\approx 16$ arcseconds) and the gaussian
shape of the nebula, whereas (b) this detail is lost in the IPC, which tracts the
surface brightness beyond $10^3$ arcseconds. Both profiles are binned around the
centroid of light in the phase-averaged image of each instrument.}
\end{figure}

This brings us to the IPC data. We gain with the IPC in two ways: with its wide
field of view ($\approx 30$ arcminutes), and with its lower background. Both of
these features allow us to track the SBP of the Crab to very large radii. In
exactly the same way as the HRI SBP, this data is presented in Figure 1b. Again,
three things are apparent: (1) due to this instrument's broad spatial
resolution, the structure of the nebula is entirely lost, (2) the flair in the
SBP is real, extending in roughly power-law form beyond $10^3$ arcseconds, and
(3) there is no hint of a shell, which would appear as a shelf-like extension of
the SBP. Obviously, we could hide a shell under this profile if it were dim and/or
small enough that it didn't disturb the power-law tail on the measured SBP. In
this way, we could place an upper limit on the luminosity of a shell from a shock
wave surrounding the Crab. We will return to this later.

\section{The Point Response Function}

In order to address the shapes observed for the SBPs, we need to consider carefully
the contributions of the point response function (PRF), which is due to the point
response of the mirror and the spatial resolution of each detector. Since the HRI
has extremely fine (0.5 arcsecond) spatial resolution, its PRF is the point
response of the mirrors. The PRF of the IPC (Fig.~2) is much more complicated due
to its broad spatial resolution. Conceptually, however, we expect the IPC's PRF to
be the convolution of the mirror response (which is measured by the HRI), and its
own intrinsic spatial resolution.

\begin{figure}[!ht]
\plotfiddle{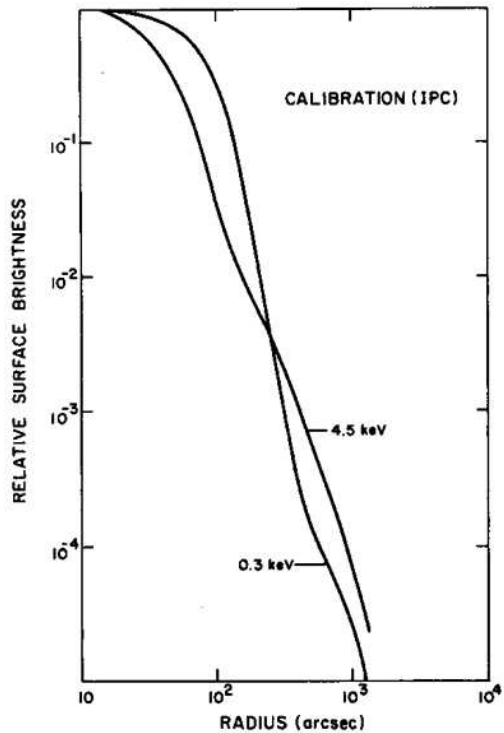}{3.96in}{0}{100}{100}{-108}{0}
\caption{The PRF for the IPC at two extreme energies spanning the Einstein
bandpass, showing the relative contributions of mirror scattering, and the effect
of the instrument's intrinsic spatial resolution.}
\end{figure}

The PRF for both the HRI and the IPC were determined during calibration testing
prior to the observatory's launch. An analysis of this data reveals that the PRF
for each instrument contains an appreciable amount of power at large radii due to
the scattering of X-rays from surface imperfections in the mirror elements. This
effect produces nearly power-law PRFs for the HRI, and adds similar wings onto the
IPC's PRFs, which are themselves dominated by the broad, nearly gaussian, spatial
resolution of this instrument.

This explains to zeroth order the observed SBPs for the Crab Nebula (Fig.~1).
The excess surface brightness beyond $\sim 10^2$ arcseconds in the HRI profile
is due to scattering by the mirror elements of X-rays from the pulsar and nebula.
Similarly, the power-law wing on the IPC's SBP is attributable to X-rays
scattered by the mirror, while the width of the core is due as much to the
spatial resolution of the detector as to the extent of the Nebula.

\section{The Model Surface Brightness Profile}

Upon closer examination, the SBP of the Crab, as measured by the IPC, has too much
power at large radii: typically, 10\% of the power of the PRFs falls beyond 3
arcminutes, whereas this fraction is 15\% for the Crab's SBP. Note that the
$\sim 50$ arcsecond extent of the nebula, being the order of, or smaller than, the
spatial resolution of the detector, and in turn being so much smaller than the
scale of the effect we are discussing here, insignificantly impacts this
comparison. The difference of 5\% between the measured SBP and the PRF is of the
order of the halo intensity reported by others for the Crab.

A more detailed investigation involves the construction of a model profile for the
response of the IPC to the intrinsic SBP of the Crab. This is handled in the
following way. To first order, the HRI measures the intrinsic SBP of the
Crab -- the differences between these being the effect of mirror scattering and
the HRI's 0.5 arcsecond spatial resolution. A second order approximation to the
Crab's intrinsic SBP would result if we could deconvolve the HRI SBP with the
HRI's PRF. We settle with truncating the HRI profile at $10^2$ arcseconds: this
removes the power-law wing obviously associated with mirror scattering, while
neglecting the broadening of the profile due to scattering interior to this
radius. This approximation to the true profile suffices if we are not interested
in, nor sensitive to, the small-scale structure of the nebula when observed with
the IPC.

We construct a model IPC SBP for the Crab by convolving the IPC's PRF with the
Crab's intrinsic SBP, which is approximated in the manner just described as the
truncated HRI SBP. The PRF we use is a linear combination of nine monoenergetic
calibration PRFs, weighted is such a way as to approximate the response of the IPC
to the Crab's continuous spectrum -- careful account being made of the gain of the
detector and the modification of the Crab's intrinsic spectrum by the effective
area of the mirror/detector combination. This PRF is convolved with the Crab's
`true' SBP to produce a model IPC SBP. This is compared to the measured profile in
Figure 3a: a significant difference is seen to exist between these two profiles
beyond $\sim 200$ arcseconds.

\begin{figure}[!ht]
\plotfiddle{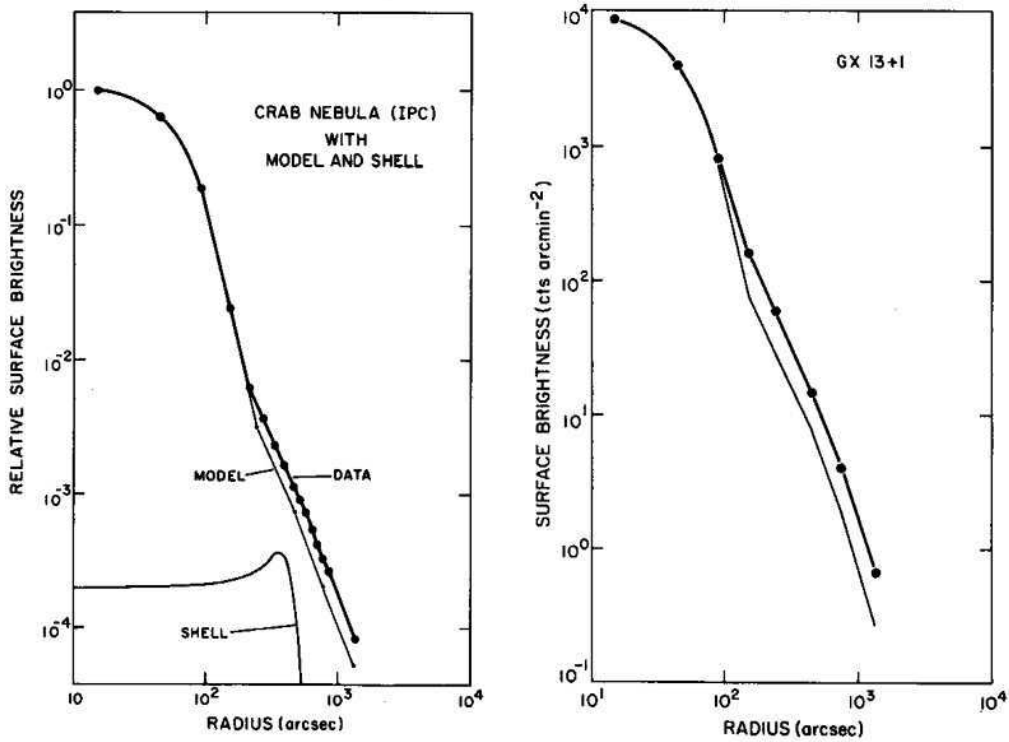}{3.99in}{0}{100}{100}{-190}{0}
\caption{(a) The SBP of the Crab Nebula (DATA) compared with the expected (MODEL)
profile. The excess surface brightness beyond $\sim 200$ arcseconds is the X-ray
halo. Also shown in the figure is the upper limit on the surface brightness of a
shell of radius 9 arcminutes. (b) The measured and model SBPs for the compact
galactic X-ray source GX13+1. The contribution to this source's halo at small
radii is due to scattering `near' the source -- such contributions to the Crab's
halo are absent because of the Crab's inclination to the galactic plane, which
brings it above the galaxy's dust layer.}
\end{figure}

\section{The X-ray Halo}

The difference between the measured and model profiles reveals the size, shape, and
intensity of the excess emission beyond the borders of the nebula. The magnitude of
this excess, defined as the fractional difference between these two profiles is
$\approx 5\%$. While this figure agrees with the magnitude of the excess emission
reported by other observers, we are able for the first time to choose
unambiguously between the two theories advanced to explain it. In particular,
this emission has the character of a halo produced by the scattering of X-rays
from interstellar grains and not that of a shock. A theoretical explanation as to
why the former interpretation is correct would take us too far afield; the
interested reader is left to pursue this topic in the literature (Overbeck 1965;
Hayakawa 1970; Mauche, et al.\ 1985). We simply assert two things: (1) the halo
around the Crab Nebula has `exactly' the same shape as the halo around the
compact galactic X-ray source GX13+1, shown in Figure 3b, and (2) the halo around
GX13+1 is due to scattering from dust.

GX13+1 is but one of the compact sources for which we have detected halos. The
strength of these have been found to correlate well with two quantities which
measure the column density of grains to a given source: (1) the amount of visual
extinction, and (2) the distance through the galaxy's dust layer. The data points
for the Crab are shown plotted amongst the points for the compact sources in Figure
4: the Crab's relatively weak halo is consistent with its extinction as well as
with its location with respect to the galactic plane.

\begin{figure}[!ht]
\plotfiddle{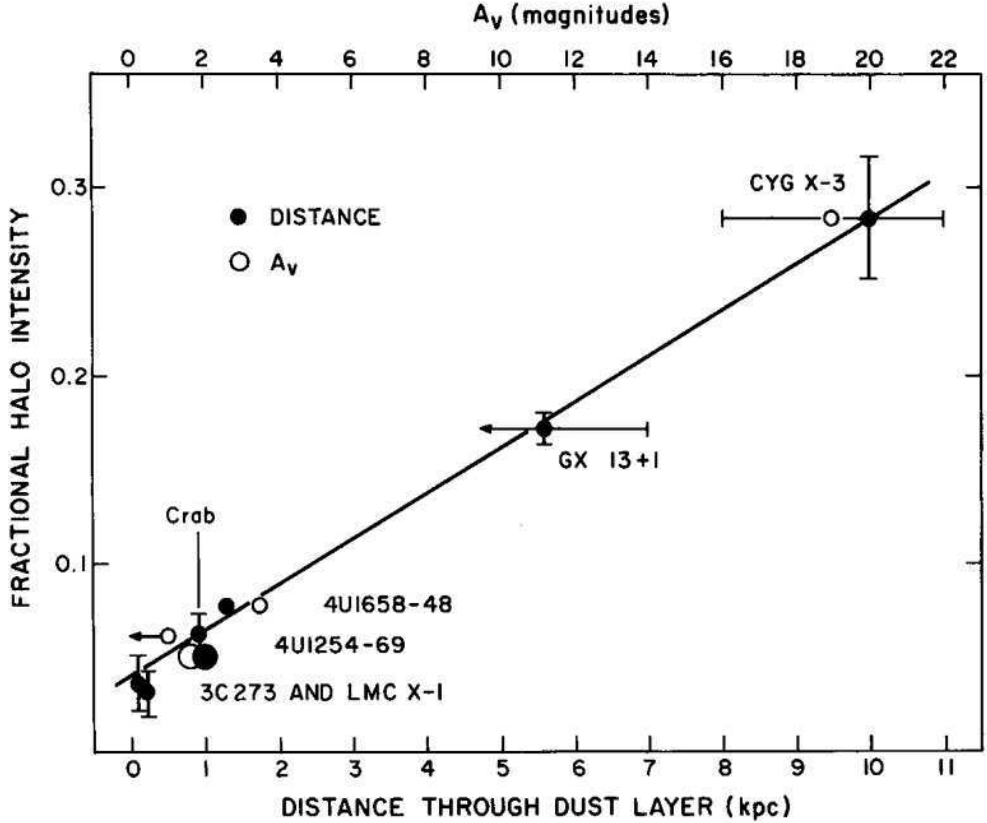}{4.39in}{0}{100}{100}{-190}{0}
\caption{The fractional halo intensity vs.\ (1) the number of magnitudes of
visual extinction for each source and (2) the distance through a dust layer of
half-thickness 0.1 kpc: for those sources out of the plane this distance is
$\rm D(kpc) = 0.1/sin(b^{II})$, whereas for those sources in the plane this
distance is equal to the entire length to the source. The line is simply fit by
eye.}
\end{figure}

\section{The Shell}

These two observations -- the similarity between Figures 3a and 3b, and the
correlation in Figure 4 -- unmistakably identifies the mechanism for the Crab's
halo as scattering from dust. However, we need to return to what the halo is not:
emission from a shell associated with a shock wave. An upper limit on the
luminosity of such a shell is dependent on its angular size, due to the obscuring
effects of mirror and dust scattering on the IPC SBP, in such a way that the upper
limit is smaller (larger) for a shell of larger (smaller) size. We choose to
calculate the upper limit for a size of 9 arcminutes, consistent with the
prediction of a Sedov solution with nominal parameters, as well as with the (age
and distance) scaled size of Cas A. Our estimate of the maximum surface brightness
for a shell of this size is shown in Figure 3a. The associated 0.5--4 keV
luminosity is about 1\% of the Crab's, or about $2\times 10^{35}~\rm erg~s^{-1}$.
Significantly, this upper limit is a factor of $\sim 6$ larger than the
luminosity (though only a factor of three in central surface brightness) of the
SN 1006 remnant, which is of approximately the same age as the Crab.
Consequently, we can not place a strong upper limit on the shell's luminosity.
The important result of this paper is however that at `soft' X-ray energies the
Crab's dust-scattering halo completely obscures its putative shell. Due to the
$\rm E^{-1}$ dependence in the size of the halo, and the $\rm E^{-2}$ dependence
in the halo intensity, the prospects for detecting the shell would be much better
at higher energies.

\acknowledgments

The authors wish to acknowledge the financial support of NASA contract NAS8-30751.

\end{document}